%% file: main.tex
\documentclass[final,3p,authoryear,11pt]{elsarticle}
\input{preamble}

\begin{document} 

\title{\textbf{Market Making with Exogenous Competition}}
\tnotetext[label0]{We thank \'Alvaro Cartea and Gerardo Duran-Martin for insightful comments on earlier versions of the model.
}

\author[imperial]{Robert Boyce}
\author[warwick]{Martin Herdegen}
\author[math,omi]{Leandro Sánchez-Betancourt}

\address[imperial]{Department of Mathematics, Imperial College London}
\address[warwick]{Department of Statistics, University of Warwick}
\address[math]{Mathematical Institute, University of Oxford}
\address[omi]{Oxford-Man Institute of Quantitative Finance, University of Oxford}
\journal{SIFIN}

\begin{frontmatter}

\begin{abstract}
We study liquidity provision in the presence of exogenous competition.
We consider a `reference market maker' who monitors her inventory and the aggregated inventory of the competing market makers. We assume that the competing market makers use 
a `rule of thumb' to determine their posted depths, depending linearly on 
their inventory. By contrast, the reference market maker optimises over her posted depths, and we assume that her fill probability depends on the difference between her posted depths and the competition's depths in an exponential way. For a linear-quadratic goal functional, we show that this model admits an approximate closed-form solution. We illustrate the features of our model and compare against alternative ways of solving the problem either via an Euler scheme or state-of-the-art reinforcement learning techniques.
\vspace{0.5em}

\vspace{0.5em}
\end{abstract}


\end{frontmatter}

\input{introduction}

\section{The model \label{sec:the model}}

Our model aims to understand how a reference market maker deals with competition. More precisely, we assume that there is one reference market maker (referred to as “she” in the sequel) who takes the presence of other market makers into account for her trading decisions, and that the  competing market makers  simply follow a “rule of thumb” trading rule derived from the classical model by \cite{avellaneda2008high}. The latter model entails that approximately the optimal ask and bid depths should be linear functions of current inventory, with a negative dependence on the ask and a positive dependence on the bid side, and yield a constant bid-ask spread; see \citet[Proposition 3]{gueant2013solution} for details. Due to this linear feature, we may aggregate without loss of generality all competing market makers into one and henceforth only speak of the competing market maker (referred to as “he” in the sequel).

We fix a filtered probability space $(\Omega, \mathcal{F}, \F = \{\mathcal{F}_{t}\}_{0 \leq t \leq T}, \mathbb{P})$ where $\mathbb{P}$ is the physical measure and $T > 0$ denotes the time horizon. As in \cite{avellaneda2008high}, the unaffected price of the asset is given by
\begin{equation} 
    S_{t} = S_{0} + \sigma\, W_{t}, \qquad t \in [0, T],
\end{equation}
where $(W_t)_{0 \leq t \leq T}$ is an $\F$-adapted standard Brownian motion.

The reference market maker's problem is to choose optimal ask and bid depths $\boldsymbol{\delta} = (\delta^{a}, \delta^{b})$ at which to post. We assume them to be $\F$-predictable 
and integrable in the  sense that $\E[\int_0^T (|\delta^a_t| + |\delta^b_t|) \dd t] < \infty$.

Given a control $\boldsymbol{\delta}$, the reference market maker posts ask and bid quotes given by
\begin{equation}
\begin{split}
    &S^{\boldsymbol{\delta},a}_{t} = S_{t} + \delta^{a}_{t}, \\
    &S^{\boldsymbol{\delta},b}_{t} = S_{t} - \delta^{b}_{t},
\end{split} \qquad  t \in [0, T].
\end{equation}
Given that the competing market maker follows a “rule of thumb” trading rule, the reference market maker indirectly influences his ask and bid depths $\boldsymbol{\tilde \delta} = (\tilde \delta^{a}, \tilde \delta^{b})$, the details of which will be specified below. The corresponding ask and bid quotes are given by
\begin{equation}
\begin{split}
    &\tilde{S}^{\boldsymbol{\delta},a}_{t} = S_{t} + \tilde \delta^{a}_{t}, \\
    &\tilde{S}^{\boldsymbol{\delta},b}_{t} = S_{t} - \tilde \delta^{b}_{t},
\end{split} \qquad  t \in [0, T].
\end{equation}
For future reference, we denote the mid-price of the competing market maker by
\begin{equation}
    \tilde{S}^{\boldsymbol{\delta}}_{t} = \frac{1}{2}\left(\tilde{S}^{\boldsymbol{\delta},a}_{t} + \tilde{S}^{\boldsymbol{\delta},b}_{t}\right), \qquad t \in [0, T].
\end{equation}
As in \cite{avellaneda2008high}, liquidity taking buy and sell orders arriving to the limit order book (LOB) are modelled by $\F$-adapted Poisson processes $(M^{a}_{t})_{0\leq t \leq T}$ and $(M^{b}_{t})_{0\leq t \leq T}$ with constant intensities $\lambda^{a} > 0$ and $\lambda^{b} > 0$. 

The buy and sell market orders that are filled by the reference market maker's limit orders are denoted by the point processes $(N^{\boldsymbol{\delta},a}_{t})_{0\leq t \leq T}$ and $(N^{\boldsymbol{\delta},b}_{t})_{0\leq t \leq T}$; the ones that are filled by the competing market maker's limit orders are denoted by the point processes $(\tilde N^{\boldsymbol{\delta},a}_{t})_{0\leq t \leq T}$ and $(\tilde N^{\boldsymbol{\delta},b}_{t})_{0\leq t \leq T}$.

The cumulative inventory of the reference market maker is given by 
\begin{equation}
Q^{\boldsymbol{\delta}}_t = N^{\boldsymbol{\delta},b}_t - N^{\boldsymbol{\delta},a}_t, \qquad t \in [0, T],
\end{equation}
whereas the cumulative inventory of the competing market maker is given by
\begin{equation}
 \tilde Q^{\boldsymbol{\delta}}_{t} = \tilde N^{\boldsymbol{\delta},b}_t - \tilde N^{\boldsymbol{\delta},a}_t, \qquad t \in [0, T].
\end{equation}
As suggested by the classical Avellaneda and Stoikov model (cf.~\citet[Proposition 3]{gueant2013solution}), we  assume that the posted depths $\boldsymbol{\tilde \delta} = (\tilde \delta^a, \tilde \delta^b)$ 
of the competing market maker depend on $\tilde Q^{\boldsymbol{\delta}}$ in a linear way and yield a constant bid-ask spread. However, we do not assume that $\tilde Q^{\boldsymbol{\delta}}$ can explain the posted depths of the competing market maker completely so we add a noise term. More precisely, we assume that 
\begin{equation} \label{eq:linear_pooled_depths}
    \tilde{\delta}^{a}_{t} = \tilde a - \beta\,  \tilde Q^{\boldsymbol{\delta}}_{t^-} - Z_{t}  \qquad \text{and} \qquad \tilde{\delta}^{b}_{t} = \tilde b + \beta \, \tilde Q^{\boldsymbol{\delta}}_{t^-} + Z_{t} ,
\end{equation}
where $\tilde a, \tilde b > 0$ denote the base ask and bid levels of the competing market maker, $\beta > 0$ is a constant of proportionality, and $(Z_{t})_{0\leq t \leq T}$ denotes a noise term satisfying $Z_{t}=\sigma_{Z}W^Z_{t}$, where $\sigma_{Z}> 0$ and $(W^Z_{t})_{0\leq t \leq T}$ is a standard Brownian motion that is independent of $W$.

Similar to the assumptions made in the extant literature on the Avellaneda and Stoikov model (cf.~\cite{gueant2013solution,cartea2015book}), we assume that the reference market maker has inventory constraints at $\underline{q},\overline{q}\in\mathbb{Z}$ with $\underline{q}<0<\overline{q}$ such that (i) if $Q^{\bm{\delta}}_{t^-} = \overline{q}$, she does not post a bid quote at time $t$, and (ii) if $Q^{\bm{\delta}}_{t^-} = \underline{q}$, she does not post an ask quote at time $t$.

It remains to explain how the choice of the reference market maker's depths $\boldsymbol{\delta} = (\delta^{a}, \delta^{b})$ influences the intensities $\Lambda^a$ and $\Lambda^b$ of $N^{\boldsymbol{\delta},a}$ and $N^{\boldsymbol{\delta},b}$ respectively. Of course, this in turn then also determines the intensities of $\tilde N^{\boldsymbol{\delta},a}$ and $\tilde N^{\boldsymbol{\delta},b}$.

As in the classical Avellaneda and Stoikov model, we assume exponential fill probabilities but with the difference that the reference market maker takes the depths of the competing market maker into account. Denoting by $\kappa > 0$ the rate of the exponential decay and $\iota > 0 $ the tick size, we posit that
\begin{equation}
\label{eq: fill prob ask}
\begin{split}
    \Lambda^{a} &:= \lambda^{a} \min\bigg(\exp\left(-\kappa\left(\delta^a - \tilde{\delta}^{a} + \iota \right)\right), 1\bigg), \\
    \Lambda^{b} &:= \lambda^{b} \min\bigg(\exp\left(-\kappa\left(\delta^b - \tilde{\delta}^{b} + \iota \right)\right), 1\bigg).
    \end{split}
\end{equation}
The economic interpretation is as follows: If the reference market maker posts one tick more generous than the competition, her limit order will be filled with probability one if a liquidity taking order  arrives on the appropriate side. Otherwise, if she posts less generously, her order is filled with an exponentially decaying probability depending on how less generous than the competition she is. Note that the classical Avellaneda and Stoikov model corresponds to the case that $\tilde \delta^a = 0 = \tilde \delta^b$ (and $\iota =0$). 

From a mathematical perspective, the parameter $\iota$ in \eqref{eq: fill prob ask} can be `absorbed' into $\tilde a$ and $\tilde b$ in \eqref{eq:linear_pooled_depths}. Thus, for mathematical convenience, from this point forward, we omit the parameter $\iota$ from the fill probabilities, and work under the assumption that $\tilde a, \tilde b$ have been adjusted to include the tick parameter, so that $\tilde{\delta}^{a}$ and $\tilde{\delta}^{b}$ denote depths that are one tick more generous than the ones of the competing market maker.

The goal functional of the reference market maker is given by
\begin{align}
J(\boldsymbol{\delta}) &= \E\Bigg[
    X_{T}^{\boldsymbol{\delta}} + Q^{\boldsymbol{\delta}}_{T}\,\tilde{S}^{\boldsymbol{\delta}}_{T}-\gamma \,(Q^{\boldsymbol{\delta}}_{T})^{2} - \phi \int_{0}^{T} (Q^{\boldsymbol{\delta}}_{r})^{2} \diff r 
    \Bigg] \\
    &= \mathbb{E}\Bigg[
    X_{T}^{\boldsymbol{\delta}} + Q^{\boldsymbol{\delta}}_{T}\bigg(S_{T} +\frac{\tilde a - \tilde b
}{2}- \beta \tilde Q^{\boldsymbol{\delta}}_{T}  -Z_T \bigg) - \gamma (Q^{\boldsymbol{\delta}}_{T})^2
    - \phi \int_{0}^{T} (Q^{\boldsymbol{\delta}}_{r})^{2} \diff r
    \Bigg]\,.
\end{align}
Here, $(X^{\boldsymbol{\delta}}_t)_{0 \leq t \leq T}$ denotes the cash process of the reference market maker, given by
\begin{equation}
X^{\boldsymbol{\delta}}_t = X_0 + \int_{(0, t]} S^{\boldsymbol{\delta},a}_r \dd N^{\boldsymbol{\delta},a}_{r} - \int_{(0, t]} S^{\boldsymbol{\delta},b}_r \dd N^{\boldsymbol{\delta},b}_{r} = X_0 + \int_{(0, t]} (S_r + \delta^a_r) \dd N^{\boldsymbol{\delta},a}_{r} - \int_{(0, t]} (S_r -  \delta^b_r) \dd N^{\boldsymbol{\delta},b}_{r},
\end{equation}
where $\gamma$ denotes the parameter for the terminal liquidation penalty, and $\phi$ denotes the parameter of running inventory aversion. Note that the reference market maker has to use the mid price $\tilde{S}^{\boldsymbol{\delta}}_T$ of the competing market maker for terminal liquidation. The above linear-quadratic goal function is widely-used for tractability; see, e.g.,  \cite{cartea2015book,gueant2016financial}.

\section{The optimal strategy}\label{sec:the optimal strategy}
The reference market maker seeks to maximise the value function
\begin{equation}
    u(t, s, x, q, \tilde q, z)= \underset{\boldsymbol{\delta} \in \cA_t}{\sup}\,\, u^{\boldsymbol{\delta}}(t, s, x, q, \tilde q, z),
\end{equation}
where $\cA_t$ denotes the set of all $\F$-predictable processes $\boldsymbol{\delta} = (\delta^a_u, \delta^b_u)_{t \leq u \leq T}$ that are integrable in the sense that $\E[\int_t^T (|\delta^a_u| + |\delta^b_u|) \dd u] < \infty$ and
\begin{equation} 
\begin{split}
    u^{\boldsymbol{\delta}}(\boldsymbol{y}) 
    &= \mathbb{E}\Bigg[
    X_{T}^{\boldsymbol{y},\boldsymbol{\delta}} + Q^{\boldsymbol{y},\boldsymbol{\delta}}_{T}\bigg(S^{\boldsymbol{y}}_{T} +\frac{\tilde a - \tilde b
}{2}- \beta \tilde Q^{\boldsymbol{y},\boldsymbol{\delta}}_{T}  -Z^{\boldsymbol{y}}_T \bigg) - \gamma (Q^{\boldsymbol{y},\boldsymbol{\delta}}_{T})^2
    - \phi \int_{t}^{T} (Q^{\boldsymbol{y},\boldsymbol{\delta}}_{r})^{2} \diff r
    \Bigg]\,, 
\end{split}
\end{equation}
where $\boldsymbol{y} = (t, s, x, q, \tilde q, z)$ and $(X^{\boldsymbol{y},\boldsymbol{\delta}}_u)_{t \leq u \leq T}$, $(Q^{\boldsymbol{y},\boldsymbol{\delta}}_u)_{t \leq u \leq T}$, $(\tilde Q^{\boldsymbol{y},\boldsymbol{\delta}}_u)_{t \leq u \leq T}$, $(S^{\boldsymbol{y}}_u)_{t \leq u \leq T}$, and $(Z^{\boldsymbol{y}}_u)_{t \leq u \leq T}$ denote the processes $X^{\boldsymbol{\delta}}$, $Q^{\boldsymbol{\delta}}$, $\tilde Q^{\boldsymbol{\delta}}$, $S$, and $Z$ restarted at time $t$ from $x, q, \tilde q, s$, and $ z$, respectively.

Using the dynamic programming approach to stochastic control, we find that the value function satisfies the Hamilton-Jacobi-Bellman (HJB) equation
\begin{align} 
    0 &= \frac{\partial u}{\partial t}(t, s, x, q, \tilde q, z)
    + \frac{\sigma^{2}}{2} \frac{\partial u}{\partial s^{2}}(t, s, x, q, \tilde q, z) + \frac{\sigma^{2}_{Z}}{2} \frac{\partial^{2} u}{\partial z^{2}}(t, s, x, q, \tilde{q}, z)  - \phi q^{2} \\
    &\quad + \lambda^{a}\bigg(
     u(t, s, x, q, \tilde q -1, z) 
    - u(t, s, x, q, \tilde q, z) \bigg) + \lambda^{b}\bigg(
     u(t, s, x, q, \tilde q +1, z) 
    - u(t, s, x, q, \tilde q, z) \bigg) \\
    &\quad +\underset{\delta^{a}}{\sup}\bigg(\lambda^a \min(e^{-\kappa(\delta^a - \tilde a + \beta \tilde q + z)},1)\big(u(t, s, x + s+\delta^{a}, q-1, \tilde q, z) 
    -u(t, s, x, q, \tilde q -1, z) \big)\bigg)\mathbbm{1}_{\{q>\underline{q}\}}  \\
    &\quad + \underset{\delta^{b}}{\sup}\bigg(\lambda^b \min(e^{-\kappa(\delta^b - \tilde b - \beta \tilde q - z)},1)\big(u(t, s, x - s+\delta^{b}, q+1, \tilde q, z) 
    - u(t, s, x, q, \tilde q+1, z) \big) \bigg)\mathbbm{1}_{\{q<\overline{q}\}}\,,
     \\
     &\quad u(T, s, x, q, \tilde q, z) = x + q\,s + \frac{\tilde a - \tilde b
    }{2} q -\gamma\,q^{2} - \beta \,\tilde q \,q\ -z\, q.
\label{HJB}
\end{align}
The terms in this HJB equation have an intuitive interpretation. The $u(t, s, x, q, \tilde{q}-1, z)  - u(t, s, x, q, \tilde{q}, z)$ and $u(t, s, x, q, \tilde{q}+1, z) - u(t, s, x, q, \tilde{q}, z)$ terms represent the difference between the value function in the case that a market order is filled by the competing market maker and the value function in the case that no market order arrives. This difference term is multiplied by the intensity of market order arrivals to market. 
Similarly, the terms $u(t, s, x + s+\delta^{a}, q-1, \tilde{q}, z) - u(t, s, x, q, \tilde{q}-1, z)$ and $u(t, s, x - s+\delta^{b}, q+1, \tilde{q}, z) - u(t, s, x, q, \tilde{q}+1, z)$  represent the difference between the value function in the case that a market order is filled by the reference market maker and the value function in the case that a market order is filled by the competing market maker. This difference term is multiplied by the intensity of market order arrivals to market and the fill probability of the reference market maker's limit order. Notice that if the fill probability is one, the terms representing the competitors filling a market order cancel (as this is then impossible) and the resulting difference term represents the change in the value function when a market order arrives. 
Motivated by the linear-quadratic goal functional, we make the linear-quadratic ansatz
\begin{align}
    u(t,s,x,q, \tilde q, z) &:= x + q\,s - \frac{\beta}{2} q^2 - \beta \tilde q q -  z q + g(t,q),\\
c^a &:= \delta^a + \beta \tilde q +z - \frac{\beta}{2},\\
    c^b &:= \delta^b - \beta \tilde q -z -\frac{\beta}{2}.
\end{align}
for some function $g(t, q)$ to be determined. Here, the terms $c^{a}$ and $c^{b}$ represent the difference between the reference and the competing market maker's depths, adjusted for a constant. This yields the following equation for $g$
\begin{equation}
\begin{split} \label{eq:HJB h}
    0 &= 
    \frac{\partial g}{\partial t}(t,q) - \phi q^{2} + (\lambda^{a} -\lambda^b)\beta q \\
    &\quad +\underset{c^{a}}{\sup}\Big(\lambda^a \min(e^{-\kappa(c^a +  \frac{\beta}{2} - \tilde a )},1)\big(c^a + g(t,q-1)-g(t,q)\big)\Big)\mathbbm{1}_{\{q>\underline{q}\}} \\
    &\quad +\underset{c^{b}}{\sup}\Big(\lambda^b \min(e^{-\kappa(c^b +  \frac{\beta}{2} - \tilde b)},1)\big(c^{b}  + g(t,q+1)-g(t,q)\big)\Big)\mathbbm{1}_{\{q<\overline{q}\}}, \\
   g(T,q) &= \frac{\tilde a - \tilde b
}{2} q -(\gamma - \frac{\beta}{2})\,q^{2}.
\end{split}
\end{equation}
Optimising over $c^a$ and $c^b$ yields
\begin{equation}
   c^{*, a}(t,q) = \max\left(\hat c^{a}(t, q), \tilde a - \frac{\beta}{2}\right) \quad \text{and} \quad    c^{*, b}(t,q) = \max\left(\hat c^{b}(t, q), \tilde b - \frac{\beta}{2}\right),
\end{equation}
where
\begin{equation}
\hat{c}^{a}(t,q) = \frac{1}{\kappa} - g(t, q-1) + g(t, q) \quad \text{and} \quad   \hat{c}^{b}(t,q) = \frac{1}{\kappa} - g(t, q+1) + g(t, q).
\end{equation}
This in turn gives
\begin{equation}
\label{eq:delta star}
\begin{split} 
   \delta^{*, a}(t,q, \tilde q, z) &= \max\left(\hat \delta^{a}(t,q, \tilde q, z), \tilde a - \beta \tilde q - z \right), \\
   \delta^{*, b}(t,q, \tilde q, z) &= \max\left(\hat \delta^{b}(t,q, \tilde q, z), \tilde b + \beta \tilde q + z \right),
\end{split}
\end{equation}
where
\begin{equation} \label{eq:delta hat}
\begin{split}
\hat\delta^{a}(t,q, \tilde q, z) &= \frac{1}{\kappa} - g(t, q-1) + g(t, q) - \beta \tilde q - z + \frac{\beta}{2}, 
\\
\hat\delta^{b}(t,q, \tilde q, z) &= \frac{1}{\kappa} - g(t, q+1) + g(t, q) + \beta \tilde q + z + \frac{\beta}{2}.
\end{split}
\end{equation}
Note that $\tilde a - \beta \tilde q - z$ and $\tilde b + \beta \tilde q + z$ denote the (one tick more generous) depths of the competing market maker on the ask and bid side, respectively.

\subsection{Approximate closed-form solution}\label{subsec:approx sol}
In a similar vein to \cite{gueant2013solution}, where the authors find a solution to the classical problem in \cite{avellaneda2008high} by assuming that the unrestrained maximisers are non-negative, we assume that the unrestrained maximisers \eqref{eq:delta hat} are greater than or equal to the (one tick more generous) respective depths of the competing market maker. In this case, \eqref{eq:HJB h} simplifies to
\begin{equation}
\begin{split} \label{eq:HJB h approx}
    0 &= 
    \frac{\partial g}{\partial t}(t,q) - \phi q^{2} + (\lambda^{a} -\lambda^b)\beta q \\
    &\quad +\frac{\lambda^a e^{-1 -\kappa (\frac{\beta}{2} - \tilde a) }}{\kappa} \exp \big(- \kappa(g(t,q) - g(t,q-1))\big)\mathbbm{1}_{\{q>\underline{q}\}} \\
    &\quad +\frac{\lambda^b e^{-1 -\kappa (\frac{\beta}{2} - \tilde b) }}{\kappa} \exp \big(- \kappa(g(t,q) - g(t,q+1))\big)\mathbbm{1}_{\{q<\overline{q}\}}, \\
   g(T,q) &= \frac{\tilde a - \tilde b
}{2} q -(\gamma - \frac{\beta}{2})\,q^{2}.
\end{split}
\end{equation}
As in the solution to the classical Avellaneda and Stoikov problem found in \cite{gueant2013solution}, the substitution $\omega(t,q) = \exp(\kappa\,g(t,q))$ leads to the linear system of ODEs
\begin{equation}
\begin{split} \label{eq:matrixODE}
    0 = &
    \frac{\partial \omega}{\partial t}(t,q) +\kappa\left(- \phi q^{2} + (\lambda^{a} -\lambda^b)\beta q\right) \omega(t,q) \\
    &+\lambda^a e^{-1 -\kappa (\frac{\beta}{2} - \tilde a)}\omega(t,q-1)\mathbbm{1}_{\{q>\underline{q}\}} +\lambda^b e^{-1 -\kappa (\frac{\beta}{2} - \tilde b)}\omega(t,q+1)\mathbbm{1}_{\{q<\overline{q}\}},  \\
    \omega(T,q) &= \exp\bigg(\kappa\Big(\frac{\tilde a - \tilde b
}{2} q -(\gamma - \frac{\beta}{2})q^{2}\Big)\bigg).
\end{split}
\end{equation}

Solving the above system and arguing backwards, we obtain the following result, where we assume that $\overline{q}=-\underline{q}$ as in \cite{gueant2013solution}. The easy proof is left to the reader.

\begin{theorem} \label{thm:approx optimal strategy}
Assume that $\overline{q}=-\underline{q}$. Define the tridiaogonal matrix $(\mathbf{A}_{i, q})_{\underline q \leq i, q \leq \overline q}$ by
\begin{equation}
    \mathbf{A}_{i,q} = 
    \begin{cases}
        -\phi\,\kappa\,q^{2} + \beta\, \kappa\, (\lambda^{a}-\lambda^{
        b})\,q  & \text{if }i=q \,,\\
        \lambda^{a}\exp\left(-1 -\kappa \left(\frac{\beta}{2} - \tilde a\right)\right) & \text{if }i=q-1 \,,\\
        \lambda^{b}\exp\left(-1 -\kappa \left(\frac{\beta}{2} - \tilde b\right)\right) & \text{if }i=q+1 \,,\\
         0 &
        \text{otherwise}\,.
    \end{cases}
\end{equation}
and the column vector $(\mathbf{v}_q)_{\underline q \leq q \leq \overline q}$ by 
\begin{equation}
\mathbf{v}_q :=\exp\left(\kappa\left(\frac{\tilde a - \tilde b
}{2}\,q -\left(\gamma - \frac{\beta}{2}\right)q^{2}\right)\right).
\end{equation}
Define the function $\omega: [0, T] \times \{\underline q, \ldots, \overline q\} \to \R$ by
\begin{equation} \label{eq:boldomega}
\omega(q, t) := \left(\exp\left(\mathbf{A}(T-t)\right)\mathbf{v}\right)_q,
\end{equation}   
and the function $u: [0, T] \times (0, \infty) \times \R \times \{\underline q, \ldots, \overline q \} \times \Z \times \R  \to \R$ by
\begin{equation}
u(t, s, x, q,\tilde q, z) = x + q\,s - \frac{\beta}{2} q^2 - \beta \tilde q q -  z q + \frac{1}{\kappa} \log (\omega(t, q)).
\end{equation}
Then $u$ solves the approximate HJB equation 
\begin{align} 
    0 &= \frac{\partial u}{\partial t}(t, s, x, q, \tilde q, z)
    + \frac{\sigma^{2}}{2} \frac{\partial u}{\partial s^{2}}(t, s, x, q, \tilde q, z) + \frac{\sigma^{2}_{Z}}{2} \frac{\partial^{2} u}{\partial z^{2}}(t, s, x, q, \tilde{q}, z)  - \phi q^{2} \\
    &\quad + \lambda^{a}\bigg(
     u(t, s, x, q, \tilde q -1, z) 
    - u(t, s, x, q, \tilde q, z) \bigg) + \lambda^{b}\bigg(
     u(t, s, x, q, \tilde q +1, z) 
    - u(t, s, x, q, \tilde q, z) \bigg) \\
    &\quad +\underset{\delta^{a}}{\sup}\bigg(\lambda^a e^{-\kappa(\delta^a - \tilde a + \beta \tilde q + z)}\big(u(t, s, x + s+\delta^{a}, q-1, \tilde q, z) 
    -u(t, s, x, q, \tilde q -1, z) \big)\bigg)\mathbbm{1}_{\{q>\underline{q}\}}  \\
    &\quad + \underset{\delta^{b}}{\sup}\bigg(\lambda^b e^{-\kappa(\delta^b - \tilde b - \beta \tilde q - z)}\big(u(t, s, x - s+\delta^{b}, q+1, \tilde q, z) 
    - u(t, s, x, q, \tilde q+1, z) \big) \bigg)\mathbbm{1}_{\{q<\overline{q}\}}
     \,, \\
     &\quad u(T, s, x, q, \tilde q, z) = x + q\,s + \frac{\tilde a - \tilde b
}{2} q -\gamma\,q^{2} - \beta \,\tilde q \,q\ -z\, q.
\label{HJB approx}
\end{align}
Moreover, the corresponding optimisers $\hat \delta^a(t, q, \tilde q, z)$ and $\hat \delta^b(t, q, \tilde q, z)$ are independent of $x$ and $s$ and satisfy for $q \in \{\underline q+1, \ldots, \overline q\}$ and $q \in \{\underline q, \ldots, \overline q-1\}$, respectively,
\begin{equation}\label{eqns:optimal depths}
\begin{split}
\hat{\delta}^{a}(t,q, \tilde q, z) &= \frac{\beta}{2} + \frac{1}{\kappa} \left(1 + \log\left(\frac{\omega(t,q)}{\omega(t,q-1)}\right)\right) -\beta \tilde q - z, \\
\hat{\delta}^{b}(t,q, \tilde q, z) &= \frac{\beta}{2} + \frac{1}{\kappa} \left(1 + \log\left(\frac{\omega(t,q)}{\omega(t, q+1)}\right)\right) +\beta \tilde q + z. \\
\end{split}
\end{equation}

\end{theorem}

Let us briefly comment on how to apply Theorem \ref{thm:approx optimal strategy} in practice. Set
\begin{equation}
\label{eq:delta star star}
\begin{split} 
   \delta^{**, a}(t,q, \tilde q, z) &:= \max\left(\hat \delta^{a}(t,q, \tilde q, z), \tilde a - \beta \tilde q - z \right), \\
   \delta^{**, b}(t,q, \tilde q, z) &:= \max\left(\hat \delta^{b}(t,q, \tilde q, z), \tilde b + \beta \tilde q + z \right),
\end{split}
\end{equation}
where $\hat{\delta}^{a}$ and $\hat{\delta}^{b}$ are as in \eqref{eqns:optimal depths}. Note that $\delta^{**, a}$ and $\delta^{**, b}$ are in general different from $\delta^{*, a}$ and $\delta^{*, b}$ in \eqref{eq:delta star} since the solution $u$ to the approximate HJB equation \eqref{HJB approx} does in general not solve the true HJB equation \eqref{HJB}. Notwithstanding, both solutions are numerically often very close or even identical (if the maximums in \eqref{eq:delta star star} are given by $\hat{\delta}^{a}$ and $\hat{\delta}^{b}$). Therefore, working with \eqref{eq:delta star star} leads to good results. In the simulation results below, we will always use \eqref{eq:delta star star} to compute the optimal depths of the reference market maker.

\section{Comparative statics and numerical results} \label{sec:numerical results}
We proceed to perform  comparative statics on the approximate closed-form solution of Theorem \ref{thm:approx optimal strategy}. The model parameters of the unaffected price process are $S_0 = 100$ and $\sigma = 1$. We take the time horizon to be $T=1$,  the baseline intensities are $\lambda^a = \lambda^b = 10$, the inventory boundaries are at $\underline{q} = -10$ and $\overline{q} = 10$, the base ask and bid  levels of the competing market maker are $\tilde{a} = \tilde{b} = 0.1$,  the constant of proportionality is $\beta = 0.05$, and the rate of exponential decay is $\kappa = 2$. The tick size (which is accounted for in $\tilde{a}, \tilde{b}$) is $\iota = 0.01$, and the penalty parameters are $\phi = 0.1$ and $\gamma = 0.03$.

\begin{figure}[H]
    \centering
    \includegraphics[width=0.44\textwidth]{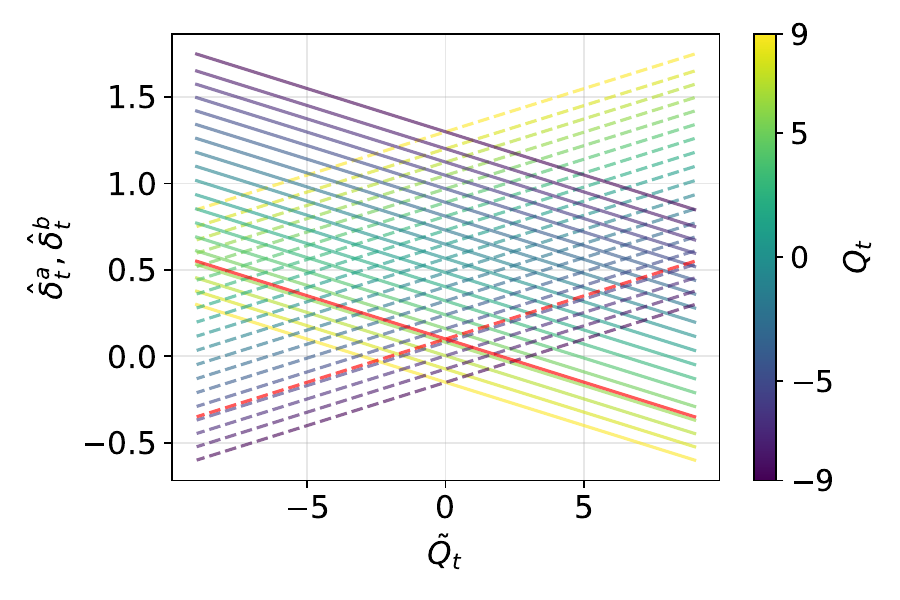}
    \includegraphics[width=0.44\textwidth]{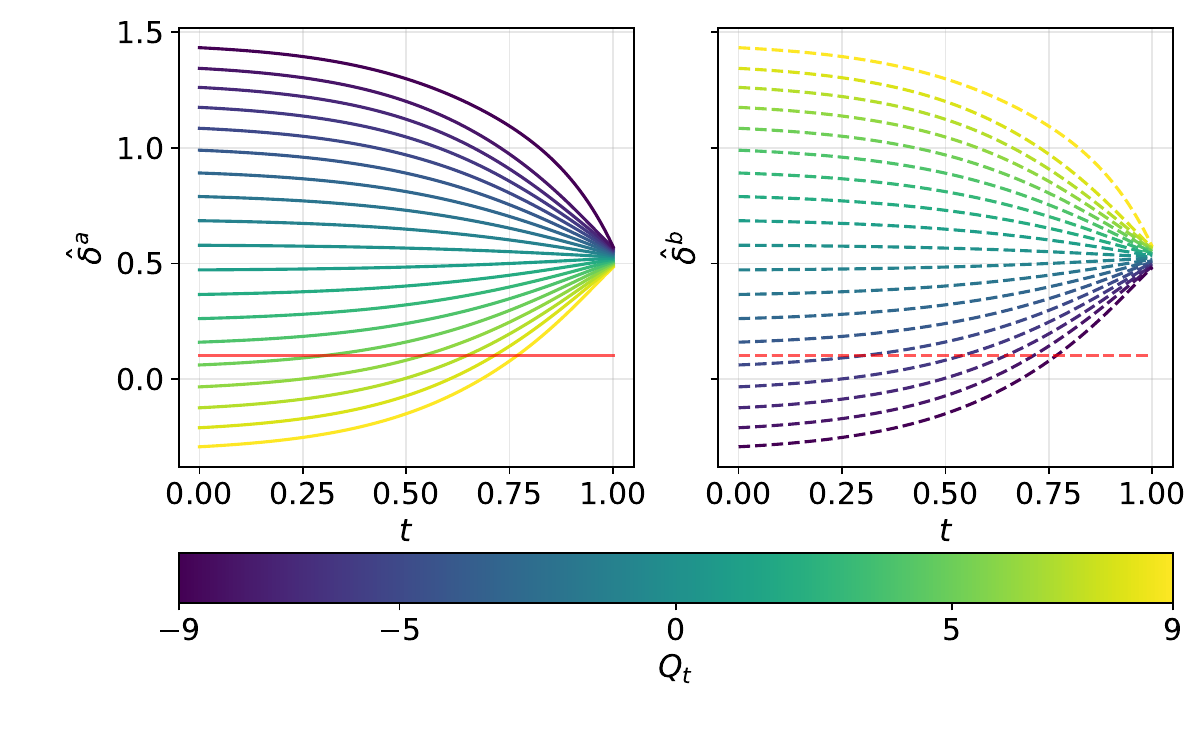}
    \caption{Left panel: Optimised ask depths $\hat\delta^{a}$ (solid lines) and bid depths $\hat\delta^{b}$ (dashed lines) at time $t=0.5$ (and for $Z = 0$) as a function of the inventory of the competing market maker ($\tilde Q_t$ in the  $x$-axis) and the inventory of the reference market maker ($Q_t$ in the colour bar). The red lines represent the depth of the competition where truncation would happen for $\delta^{**,a}$ and $\delta^{**,b}$. Right panel: Optimised depths $\hat\delta^{a}$ and $\hat\delta^{b}$ for $\tilde{Q} = 0$ (and $Z = 0$) as a function of time ($t$ in the $x$-axis) and the inventory of the reference market maker ($Q_t$ in the colour bar). The red lines show the competing market makers's depths, $\tilde{\delta}^{a}$ and $\tilde{\delta}^{b}$, including the tick size.}
    \label{fig:approxoptdepths.pdf}
\end{figure}

Figure \ref{fig:approxoptdepths.pdf} shows the optimised depths in \eqref{eqns:optimal depths} as a function of (i) the inventory $\tilde{Q}_t$ of the competing market maker and the inventory $Q_t$ of the reference market maker in the left panel, and as a function of (ii) time and the inventory the inventory $Q_t$ of the reference market maker in the right panel.

As expected, the larger the inventory $Q_t$ of the reference market maker is, the more (resp.~less) generous the ask (resp.~bid) quotes are; this is because the reference market maker has an incentive to revert her inventory to zero. As a function of the inventory of the competing market maker, the effect is the reverse: the larger the inventory $\tilde Q_t$ of the competition is, the less (resp.~more) generous the ask (resp.~bid) quotes are; this follows from the mechanics \eqref{eq:linear_pooled_depths} of how the competing market maker posts quotes based on his inventory. Note that in order to compute the ‘actual’ ask and bid depths $\delta^{**,a}$ and   $\delta^{**,b}$ in \eqref{eq:delta star star} one has to take the maximum between the plotted ask and bid depths $\hat \delta^{a}$ and   $\hat \delta^{b}$ and the bid and ask depths $\tilde \delta^{a}$ and   $\tilde \delta^{b}$ of the competition.

\begin{figure}[H]
    \centering
    \includegraphics[width=0.32\textwidth]{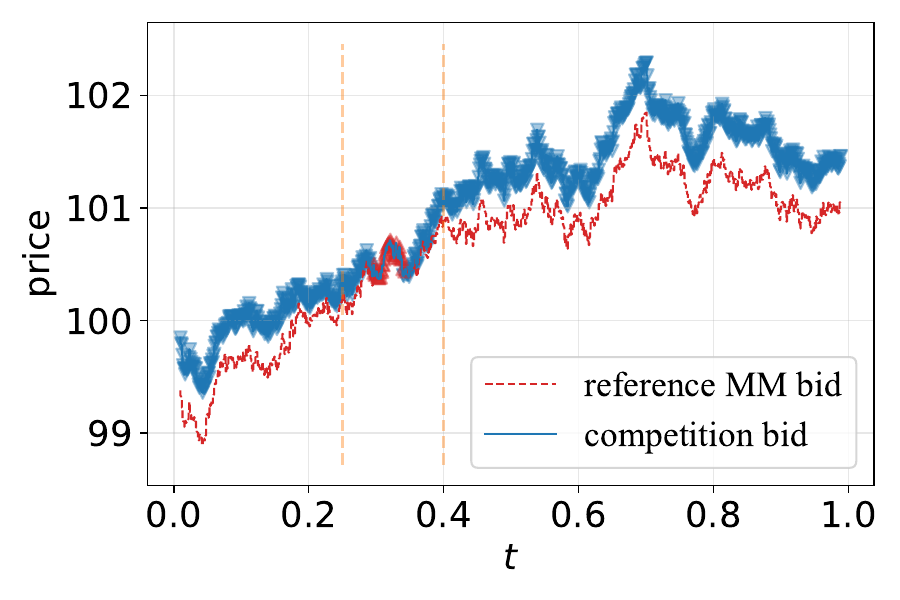}
    \includegraphics[width=0.32\textwidth]{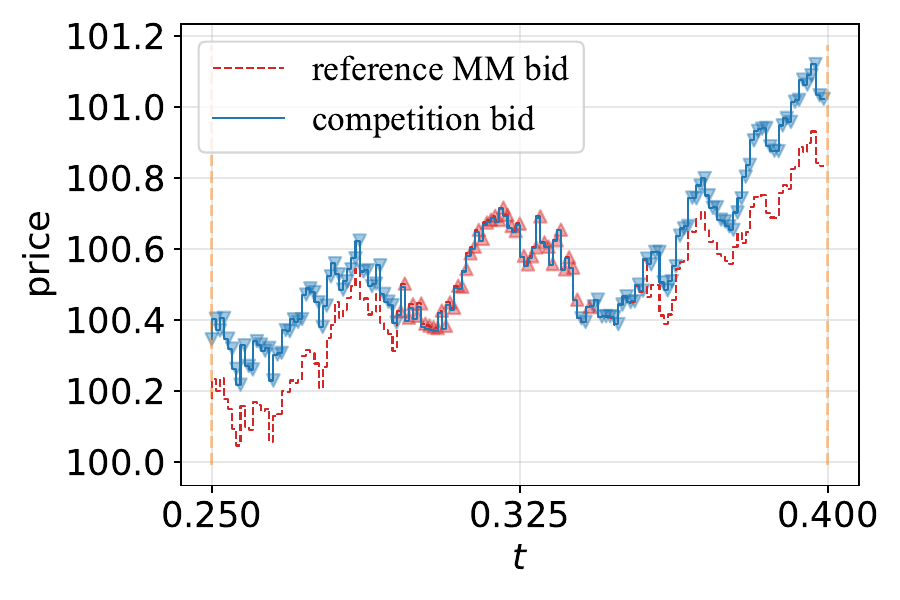}
    \includegraphics[width=0.32\textwidth]{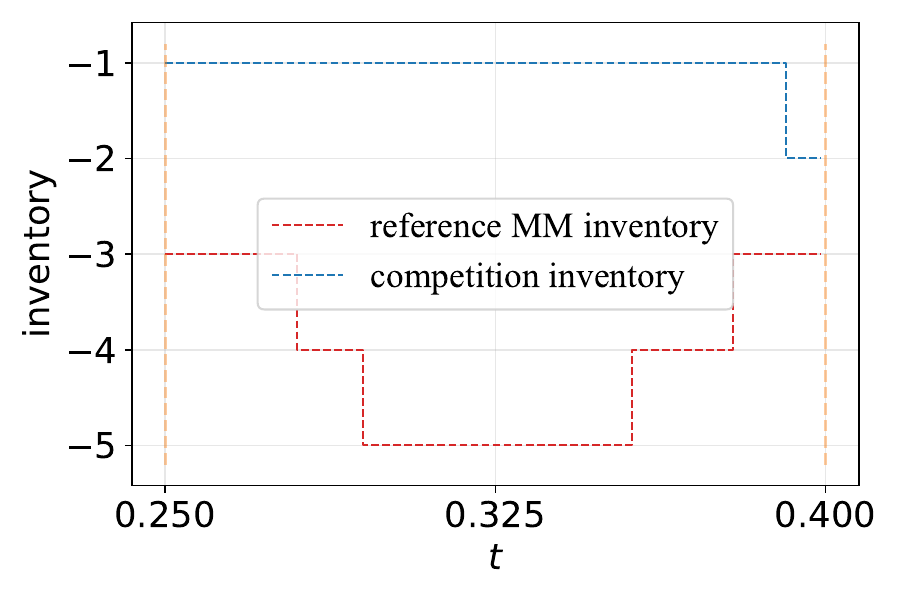}
    \caption{Simulation in which the reference market maker becomes more generous than the competition. Blue markers denote that the competition has the best quote, and red markers denote that the reference market maker has the best quote. The left panel shows the best bid quotes for the entire day. The middle panel zooms into the portion of the trading day in which the best bid quotes of the reference market maker becomes more generous than competition. The right panel shows the inventories over the zoomed period.}
\label{fig:competition_for_TOB}
\end{figure}

Next, we discretise $[0,T]$ in 1,000 timesteps and run 10,000 simulations to study  the approximated  closed-form solution  \eqref{eqns:optimal depths} in more detail. First, out of the 10,000 simulations, the reference market maker was more generous than the competition (either in the ask quotes or bid quotes) thirteen times (0.13\% of all simulations). Figure \ref{fig:competition_for_TOB} shows one of these thirteen simulations.

We observe that a large negative value for the inventory $Q_t$ forces the reference market maker to be one tick better than competition until an incoming sell market order arrives and her inventory decreases --- after this, she abandons the competition for the top of book.

\begin{figure}[H]
    \centering
    \includegraphics[width=0.8\textwidth]{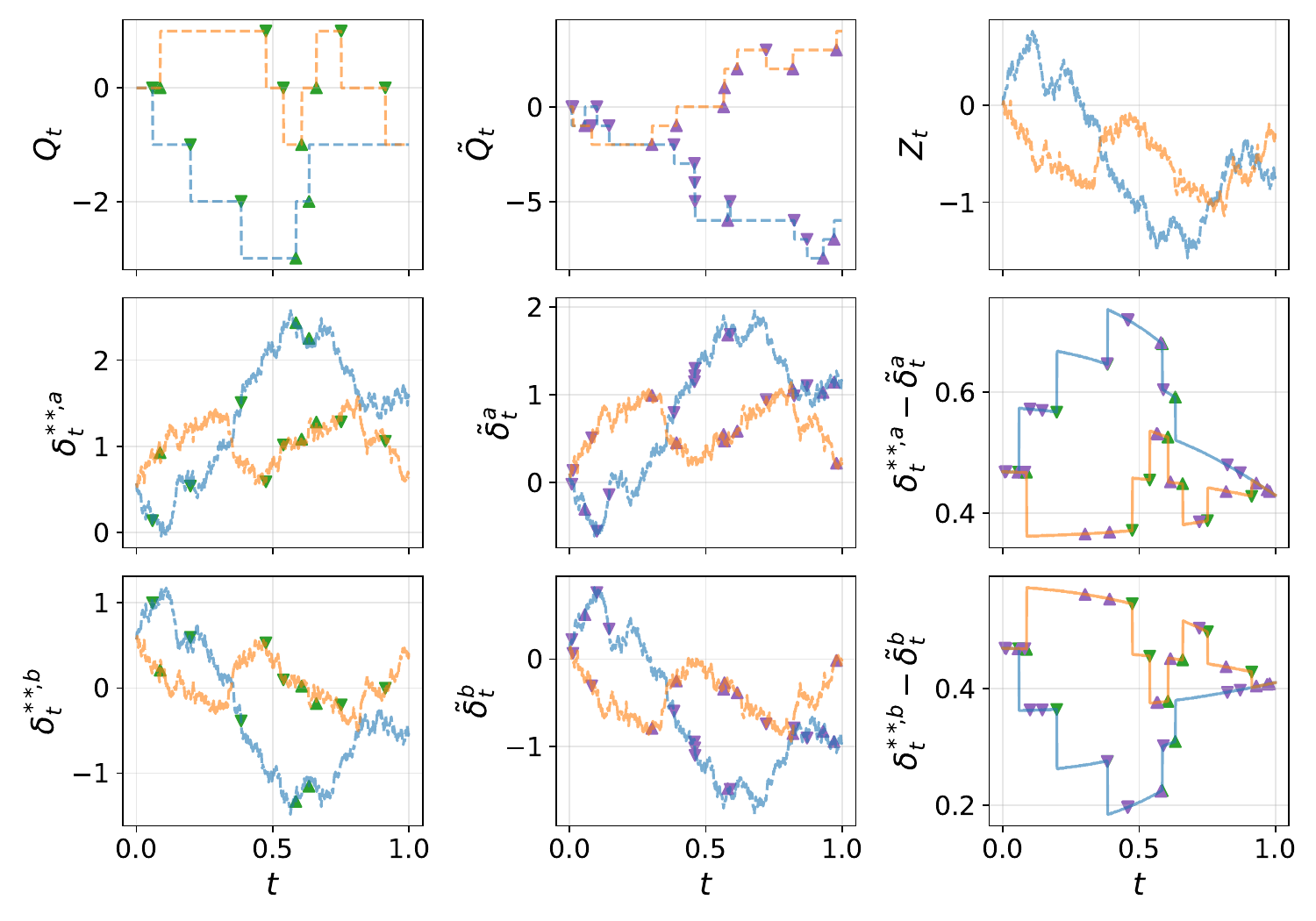}
    \caption{First row: two sample paths for the inventory processes $Q_t$ (top-left) and $\tilde{Q}_t$ (top-centre), and the noise process $Z_t$ (top-right). Second and third row: approximate closed-form depths $\delta^{**,a}_t,\delta^{**,b}_t$ (left column) in \eqref{eqns:optimal depths}, the competition depths $\tilde\delta^a_t,\tilde\delta^b_t$ (middle column), and the difference between them (right column). Filled buy trades are up-facing markers in green for the reference market maker and in purple for the competition; similarly, filled sell trades are down-facing markers in green for the reference market maker and in purple for the competition. The two sample paths are coloured in blue and orange.}
    \label{fig:q_d_deltas_tildes}
\end{figure}

Figure \ref{fig:q_d_deltas_tildes} shows two 
sample paths (one in orange and one in blue) of the main processes.

We see that the approximate closed-form depths $\delta^{**,a}_t,\delta^{**,b}_t$  in \eqref{eqns:optimal depths} follow the competition which in turn is largely influenced by the noise process $Z_t$ and their inventory $\tilde Q_t$. The most informative plots are those in the right of the middle and bottom rows. They show the difference between the quotes of the competition and the reference market maker (ask in the middle row and bid in the bottom row). As expected the gaps in the ask quotes (resp.~bid quotes) close/widen depending on whether the inventory increases/decreases (resp.~decreases/increases) and the gap closes towards the end of the trading day because the mark-to-market is done at the competition's midprice.

Finally, we study how good the approximate closed-form solution is compared to using a numerical approximation of the  solution to the original problem. For this we compare against two benchmarks: (i) Euler-scheme-type approximation of the original value function using 1,000,000 equally-spaced points in the discretisation, and (ii) reinforcement learning on the original problem using proximal policy optimisation (PPO).

\begin{figure}[H]
    \centering
    \includegraphics[width=0.4\textwidth]{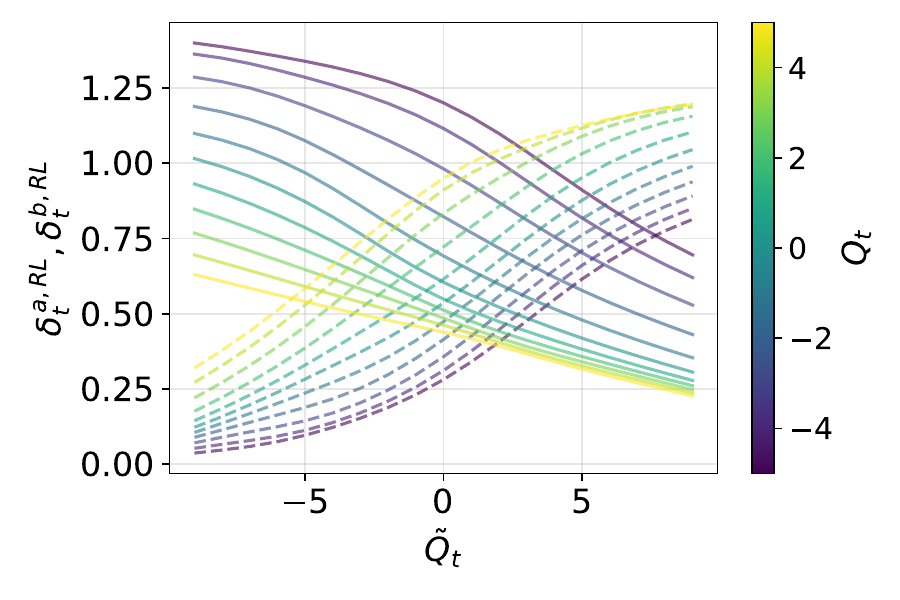}
    \includegraphics[width=0.4\textwidth]{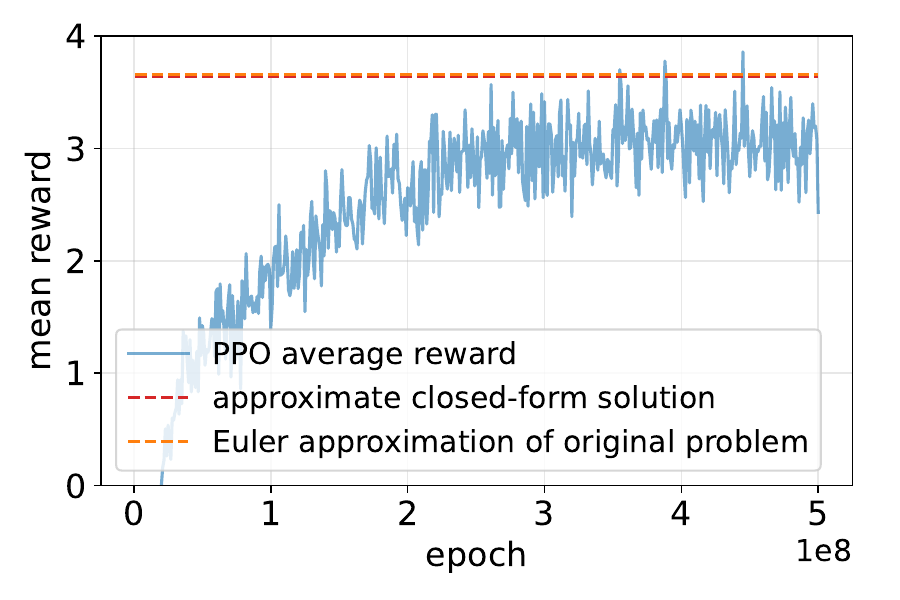}
    \caption{Left panel shows the policy learnt  by the PPO agent. Right panel compares the performance of the PPO agent against  (i) the approximate closed-form solution and the (ii) Euler-approximation of the solution to the original problem. }
    \label{fig:RL}
\end{figure}

We find that the average (with standard deviation) of the performance criterion of the reference market maker is 3.64 (2.57) following the closed-form approximate solution,  3.66 (2.56) following the Euler-scheme-type approximation  to the real solution, and 3.14 (2.76) following the reinforcement learning policy on the original problem. The 0.5\% increase when going from the approximate closed-form solution to the solution that uses the Euler-scheme approximation of the original value function is significant (according to a paired $t$-test at 99\% confidence). The under-performance of PPO might be corrected with fine-tuning of policy (or value function) parameters or architectures. We train the reinforcement learning agent over 500 million epochs and our results are publicly available.\footnote{Our code is publicly available at \url{https://github.com/leandro-sbetancourt/mm-pooled-competition} and \url{https://github.com/leandro-sbetancourt/gym-mm-pooled-competition}. We build on the gym environment `mbt\_gym' by \cite{jerome2022model}.} 
Figure \ref{fig:RL} summarises these findings.

In the left panel we see that the learnt PPO policy shares some similarities with the left panel in Figure \ref{fig:approxoptdepths.pdf}; e.g., ordering of the lines, level,  and direction.  The right panel shows the average reward through the learning process (500 million epochs in total). We see that  both (i) the approximate closed-form solution and the (ii) Euler-approximation of the solution to the original problem, are above the mean of the observed performance of the PPO agent.

\section{Conclusion}\label{sec:conclusions}
In this paper, we have considered a model where a market maker set quotes to maximise trading revenue while minimising inventory risk in the presence of competitors who adjust their quotes with the orders they fill. Orders not filled by the reference market maker will instead be filled by the competitors. 
This means the reference market maker faces a tradeoff between managing her own inventory and that of the competitors. We derive the optimal quotes in feedback form for the market maker and, using a method inspired by that of \cite{gueant2013solution}, find an approximate closed-form solution. We then numerically investigate the true solution and approximate closed-form solution, and compare the performance to that of state-of-the-art reinforcement learning.

\bibliographystyle{plainnat}
\bibliography{references}

\end{document}

%% file: preamble.tex
\usepackage[colorinlistoftodos,textsize=footnotesize]{todonotes}
\usepackage{amsfonts,amsmath,amssymb,amsthm}
\usepackage{hyperref}
\hypersetup{
	colorlinks = true, 
	urlcolor = blue, 
	linkcolor = blue, 
	citecolor = blue 
}
\usepackage{mathtools}
\usepackage[utf8]{inputenc}
\usepackage[caption=true]{subfig}
\usepackage[ruled, vlined]{algorithm2e}
\usepackage{enumitem}
\usepackage{color}
\usepackage{bm}
\mathtoolsset{showonlyrefs}
\allowdisplaybreaks 
\usepackage{graphicx}
\usepackage{bbold}
\usepackage{bbm}
\usepackage{multirow, multicol}
\usepackage{array}
\usepackage{float}
\usepackage{comment}

\newcommand{\diff}{\mathrm{d}}

\theoremstyle{plain}
\newtheorem{theorem}{Theorem}[section]

\theoremstyle{definition}

\newtheorem*{assumption*}{Standing Assumption}

\theoremstyle{remark}

\numberwithin{equation}{section}

\newcommand{\E}{\mathbb{E}}
\newcommand{\F}{\mathbb{F}}

\newcommand{\R}{\mathbb{R}}

\newcommand{\Z}{\mathbb{Z}}

\newcommand{\cA}{\mathcal{A}}

\newcommand{\dd}{\,\mathrm{d}}

%% file: introduction.tex
\section{Introduction \label{sec:introduction}}

Limit order books (LOBs)  are the main technology used in exchanges today. 
Some market participants, known as `market makers', aim to `earn the spread' by providing liquidity in the LOB.
These participants also aim to reduce their exposure to asset price fluctuations, known as `inventory risk'. This presents them with a trade-off between posting prices that provide large profits and posting prices that maximise their ability to unwind positions with limit orders. In a market where a number of market makers aim to optimise this trade-off, the order flow that is not filled by a particular `reference market maker' affects her future optimal behaviour because the corresponding market orders are filled by another market maker, who adjusts prices accordingly and in turn alters the future probabilities of the reference market maker's limit orders being filled. In this paper, we seek to study the behaviour of a reference market maker in the presence of competing market makers.

The modern mathematical finance literature in market making was initiated by \cite{avellaneda2008high} who studied optimal ask and bid prices for a market maker that seeks to control her inventory risk, using ideas from the much earlier work by \cite{ho1983dynamics}. In their framework, market orders arrive according to a Poisson process and are filled according to an exponential fill probability which depends on the distance -- known as the depth -- of the market maker's posted price from the so-called `midprice' or unaffected price. This exponential fill probability represents an abstraction of the notion of market orders `walking the book'.\footnote{There are a number of alternative formulations of the market making problem; see \cite{guilbaud2013optimal,kuhn2015optimal,lu2018order,chavez2024adaptive} for examples.}


There have been various extensions and modifications to the modelling of market order arrivals. \cite{cartea2014buylowsellhigh} model the arrivals with Hawkes processes that depend on so-called `influential' market orders. \cite{cartea2020market} introduce `signals' in the market making problem. \cite{jusselin2021optimal} considers order flow information in the context of the market making problem and also makes use of Hawkes processes. \cite{bergault2021closed} extend the closed-form approximations to a multi-asset setting. \cite{cartea2017algorithmic} solve the market making problem when there is misspecification in arrival rates, fill probabilities, and the midprice process.  
We refer to
\cite{gueant2016book} and \cite{cartea2015book} for a textbook treatment of the market making problem. 


To the best of our knowledge, the effect of a market orders not being filled by a market maker has not been previously studied.\footnote{Note, however, that the effect of missed marketable limit orders has been studied in \cite{cartea2021latency} from a liquidity taking perspective.} In principle, a market order not filled by the reference market maker affects the posted depths of the other market makers; cf.~\cite{chorida2002orderimbalance}. The Avellaneda-Stoikov model and the subsequent literature do not consider the interaction between several market makers. In particular, the fill probabilities depend only on the reference market maker without considering the impact of other market makers.\footnote{Some papers that study the presence of other market makers (but not for limit orders) include \cite{herdegen2023liquidity}, who consider market makers competing for market orders in a one-shot Nash equilibrium in a Stackelberg game, and \cite{luo2021dynamiceqpricecomp}, who study theoretical properties -- including existence and uniqueness of Nash equilibria -- in a continuous time stochastic differential game where market makers have incomplete information.}

In this work, we introduce a model in the spirit of \cite{avellaneda2008high} where unfilled market orders affect future fill probabilities because the competing market makers filling such orders adjust their quotes appropriately. We assume that the competing market makers use a ‘rule of thumb’ for their posted depths, depending in a linear way on their inventory. For this reason we can aggregate them into one. Moreover, our approach allows us to model the effect of the competitor's inventories on market conditions without resorting to the theory of stochastic differential games, making our model more tractable and allowing for an approximate closed-form solution. In our model, the reference market maker is constantly faced with a trade-off between managing her own inventory risk and managing the number of unfilled market orders in an attempt to make market conditions more favourable for herself in the future. When the intensities of market order arrivals to market differ significantly, as might be expected in a period with large changes in market returns (see e.g.~\cite{chorida2002orderimbalance}), this trade-off becomes challenging to manage.

In a similar spirit to the solution of the market making problem with inventory risk found in \cite{gueant2013solution}, and presented for a linear-quadratic objective function in \cite{cartea2015book}, we derive an approximate closed-form solution to the reference market maker's problem in our model. This is done by making an assumption about the relation between the competitors' depths and the reference market maker's depths, which usually holds for appropriate parameters. This approximation is similar to the one used in \cite{gueant2013solution} and involves the solution to a matrix ordinary differential equation (ODE). 


The remainder of the article is organised as follows. In Section \ref{sec:the model} we introduce our model and describe the optimisation problem of the reference market maker in detail.  In Section \ref{sec:the optimal strategy} we derive the optimal strategy in feedback form and describe the approximate closed-form optimal strategy in Theorem \ref{thm:approx optimal strategy}, using a method inspired by \cite{gueant2013solution}. In Section \ref{sec:numerical results} we illustrate some interesting features of our model by performing various comparative statics and also briefly compare the approximate solution to the numerical solution of the problem via an Euler scheme or a reinforcement learning agent using proximal policy optimisation (PPO).\footnote{ 
The PPO algorithm was introduced in \cite{schulman2017proximal}.
} Section \ref{sec:conclusions} concludes.